\documentclass[a4paper]{jpconf}
\usepackage{graphicx}
\begin{document}
\title{Energy dependence of fluctuations in p+p and Be+Be collisions from NA61/SHINE}

\author{Evgeny Andronov for the NA61/SHINE Collaboration}

\address{Saint Petersburg State University, ul. Ulyanovskaya 1, 198504, Petrodvorets, Saint Petersburg, Russia}

\ead{evgeny.andronov@cern.ch}

\begin{abstract}
The main aims of the NA61/SHINE experiment are to discover the critical point of strongly interacting matter and to study properties of the onset of deconfinement. For this goal a two dimensional scan of the phase diagram (T-$\mu_{B}$) is being performed at the SPS with measurements of hadron production in proton-proton, proton-nucleus and nucleus-nucleus interactions as a function of collision energy and system size. It is expected that fluctuations of various dynamical quantities will increase close to the critical point. In this contribution recent results of measurements of fluctuations in p+p and Be+Be collisions at SPS energies are presented as well as comparisons with model predictions and corresponding data of other experiments are discussed.
\end{abstract}

\section{Introduction}
The NA61/SHINE experiment~\cite{Abgrall:2014fa} is a fixed target experiment at the Super Proton Synchrotron (SPS) of the European Organization for Nuclear Research (CERN). The strong interaction programme of NA61/SHINE consists of studies of the onset of deconfinement (OD)~\cite{Gazdzicki:1998vd},~\cite{Alt:2007aa} in heavy ion collisions and search for the critical point (CP)~\cite{Fodor:2004nz} of strongly interacting matter. A two-dimensional phase diagram scan - energy versus system size - is being performed by NA61/SHINE. It includes measurements of hadron production in collisions of protons and various nuclei (p+p, Be+Be, Ar+Sc, Xe+La) at a range of beam momenta (13{\it A} - 158{\it A} GeV/c). It is expected that there will be an enhancement of fluctuations of a number of observables due to the phase transition related to the CP and the OD~\cite{Stephanov:1999zu}. Some hints of this enhancement were already observed by the NA49 experiment~\cite{Afanasiev:1999}, predecessor of NA61/SHINE, at the top SPS energy~\cite{Grebieszkow:2009jr}.

\section{Fluctuation measures}
The size of the system created in collisions of two nuclei changes significantly from event to event influencing fluctuations of measured quantities. The following classification of observables was introduced: 1) extensive quantity - proportional to the system volume in the Grand Canonical Ensemble or the number of the wounded nucleons in the Wounded Nucleon Model~\cite{Bialas:1976ed} 2) intensive quantity - independent of the system volume 3) strongly intensive quantity~\cite{Gorenstein:2011vq} - independent of the system volume and fluctuations of this volume. Strongly intensive quantities are best suited to study fluctuations in nucleus-nucleus collisions because of the unavaoidable event-to-event variations of the volume.

The most common way to characterize fluctuations of the quantity {\it A} is to measure the variance of its distribution $Var(A)$ which is an extensive quantity. The scaled variance
\begin{equation}
            \omega[A] = \frac{Var(A)}{\langle A \rangle}
\end{equation}
is intensive. For the Poisson distribution $\omega=1$. Section~\ref{mult} is devoted to this quantity applied in the analysis of multiplicity fluctuations.

As was shown in~\cite{Gorenstein:2011vq}, there are two ways to combine the first two moments of extensive observables {\it A} and {\it B} to get strongly intensive quantities:
\begin{eqnarray}
            &\Delta[A,B] = \frac{1}{C_{\Delta}} \biggl[ \langle B \rangle \omega[A] -
                        \langle A \rangle \omega[B] \biggr] \\
            &\Sigma[A,B] = \frac{1}{C_{\Sigma}} \biggl[ \langle B \rangle \omega[A] +
                        \langle A \rangle \omega[B] - 2 \bigl( \langle AB \rangle -
                        \langle A \rangle \langle B \rangle \bigr) \biggr]
\end{eqnarray}
The normalization factors $C_{\Delta}$ and $C_{\Sigma}$ are usually chosen in such a way~\cite{Gazdzicki:2013ana} that for independent particle emission $\Delta[A,B] =\Sigma[A,B] = 1$. In Section~\ref{charge} results for electric charge fluctuations are presented. In this case:
$$
            A = N_{+}, \qquad B = N_{-}, \qquad
            C_{\Delta} = \langle N_{-}\rangle -\langle N_{+}\rangle, \qquad C_{\Sigma} = \langle N_{-}\rangle +\langle N_{+}\rangle.
        $$
\subsection{NA61/SHINE multiplicity fluctuation studies}
\label{mult}
The analysis of multiplicity fluctuations was performed for inelastic p+p interactions at $\sqrt{s_{NN}}=$ 7.6, 8.7, 12.3, 17.3 GeV and for centrality selected Be+Be interactions at $\sqrt{s_{NN}}=$ 6.27, 8.73, 11.94, 16.83 GeV. Results are preliminary, include the statistical uncertainty and a first estimate of systematic uncertainty.

\begin{figure}[h]
\begin{center}
\includegraphics[width=31pc]{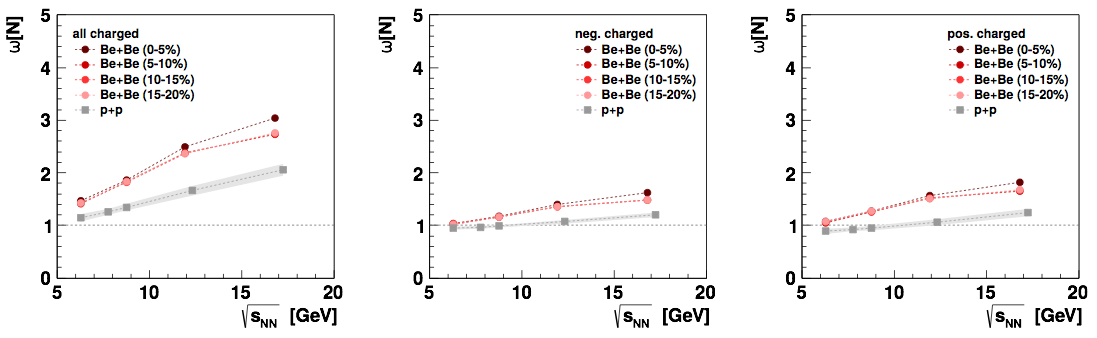}
\end{center}
\caption{\label{label1}Scaled variance of the multiplicity distribution for Be+Be collisions of several collision centralities and energies for all charged (left), negatively charged (middle) and positively charged (right) particles compared with the corresponding results from the energy scan with inelastic p+p interactions~\cite{Czopowich:2014cp}.}
\end{figure}

Figure~\ref{label1} presents results for the scaled variance of the charged hadron multiplicity distribution from Be+Be collisions and inelastic p+p interactions~\cite{Czopowich:2014cp}. $\omega$ increases with energy probably due to KNO scaling (variance of the multiplicity distribution increases faster than its mean). Larger values for Be+Be than for p+p collisions are probably due to volume fluctuations because $\omega$ is not a strongly intensive measure.

In order to measure fluctuations of particles of specific types a new identification procedure (the identity method~\cite{Gazdzicki:2011xz}) was developed. It allows to determine the second and third moments of the multiplicity distribution when the particle identification is not unique but can only be done on a statistical basis. Applying this method NA49 and NA61/SHINE determined the scaled variance of the multiplicity distribution of identified protons, kaons and pions in inelastic p+p and 3.5\% most central Pb+Pb collisions~\cite{Mackowiak-Pawlowska:2014ipa}. For $\pi$ as well as for all charged particles we observe a linear increase of $\omega$ with the collision energy and higher values of $\omega$ for the larger system. For $K$ and p these properties are probably masked by effects of baryon number conservation and strangeness conservation. 

Both for unidentified and identified charged particle fluctuations no indication of the CP is found.
\begin{figure}[h]
\begin{center}
\includegraphics[width=28pc]{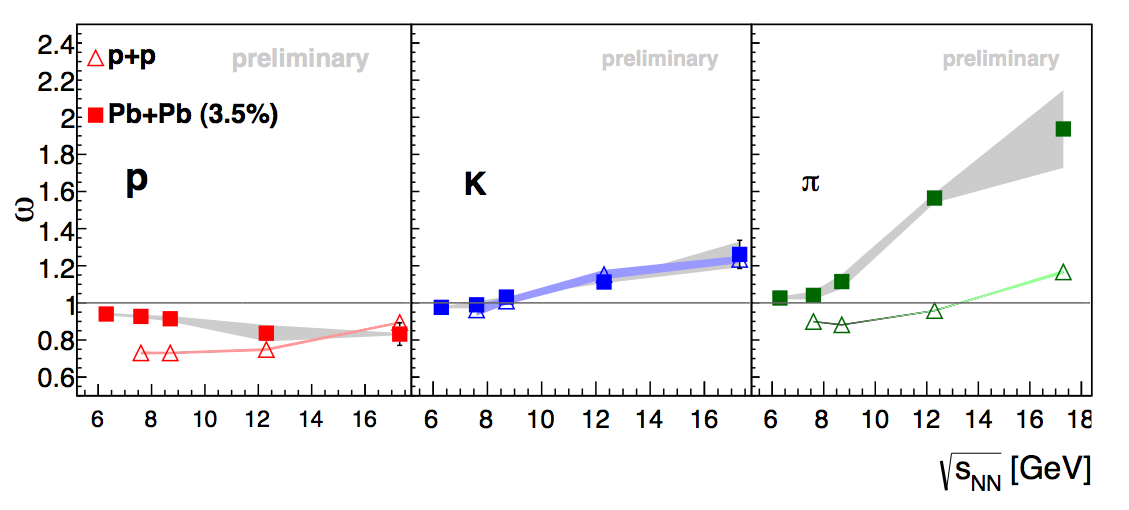}
\end{center}
\caption{\label{label2} Scaled variance $\omega$ of the multiplicity distribution of protons, kaons, and pions versus
nucleon-nucleon cms energy $\sqrt{s_{NN}}$ for the 3.5\% most central Pb+Pb collisions (NA49, full squares)
and inelastic p+p reactions (NA61, open triangles)~\cite{Mackowiak-Pawlowska:2014ipa}.}
\end{figure}
\subsection{NA61/SHINE electric charge fluctuation studies}
\label{charge}
Fluctuations of conserved charges are expected to be sensitive to phase transitions~\cite{Jeon:2000}. An analysis of fluctuations of electric charge was performed for Be+Be interactions at $\sqrt{s_{NN}}=$ 16.83 GeV for three centrality classes 0-5\%, 0-10\% and 0-20\%. Centrality was determined from the energy deposited in the NA61/SHINE forward calorimeter - the Projectile Spectator Detector (PSD). $\Delta[N_{+},N_{-}]$ and $\Sigma[N_{+},N_{-}]$ were calculated for 9 pseudorapidity intervals centred at pseudorapidity $\eta_{lab}=4.6$ in the laboratory reference frame. The width of the interval varies from 0.2 to 3.4 units of pseudorapidity. Results are preliminary and include the statistical uncertainty.

\begin{figure}[h]
\includegraphics[width=12pc]{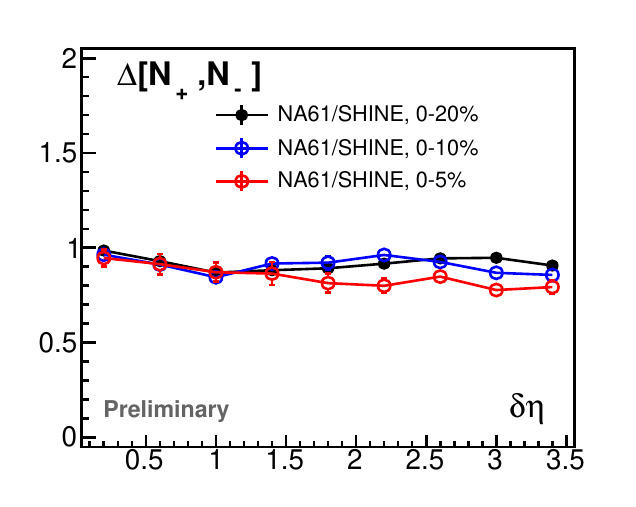}\includegraphics[width=12pc]{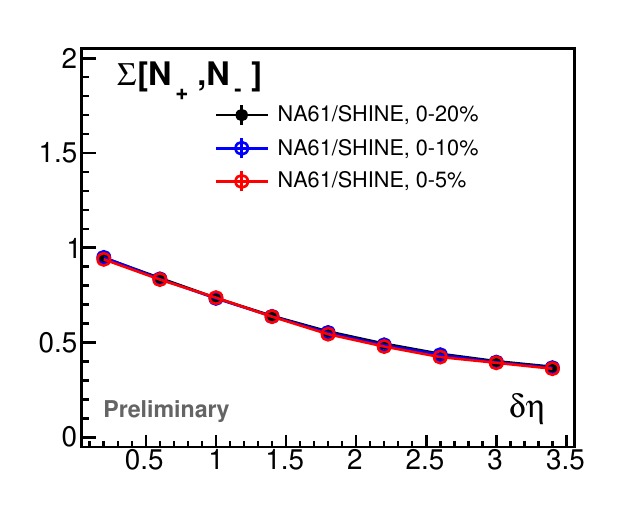}
\begin{minipage}[b]{14pc}\caption{\label{centr}$\Delta[N_{+},N_{-}]$ (left) and $\Sigma[N_{+},N_{-}]$ (right) as a function of the width of pseudorapidity interval $\delta\eta$ for 0-5\% (red), 0-10\% (blue) and 0-20\% (black) centrality classes for Be+Be interactions at $\sqrt{s_{NN}}=$ 16.83 GeV.}
\end{minipage}
\end{figure}

In Figure~\ref{centr} results for $\Delta[N_{+},N_{-}]$ and $\Sigma[N_{+},N_{-}]$ are plotted as a function of the width of the pseudorapidity interval for three centrality classes. The weak dependence on centrality in Be+Be shows that the fluctuations do not depend on the collision volume. The suppressed ($<$1) values of $\Sigma[N_{+},N_{-}]$ are probably due to global conservation of electric charge. This effect is less visible for $\Delta[N_{+},N_{-}]$.

\begin{figure}[h]
\includegraphics[width=12pc]{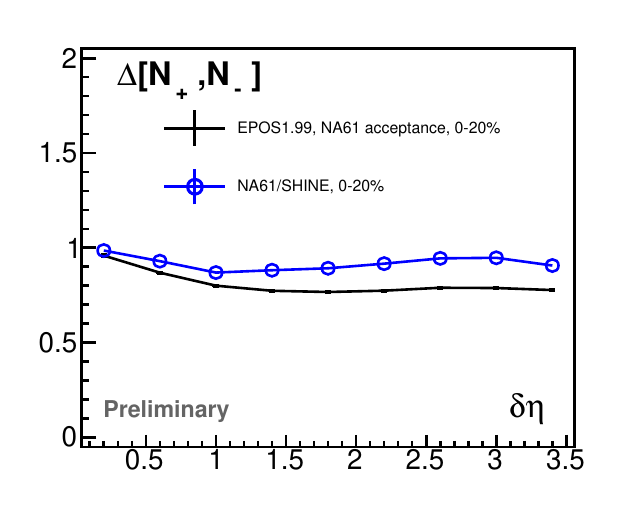}\includegraphics[width=12pc]{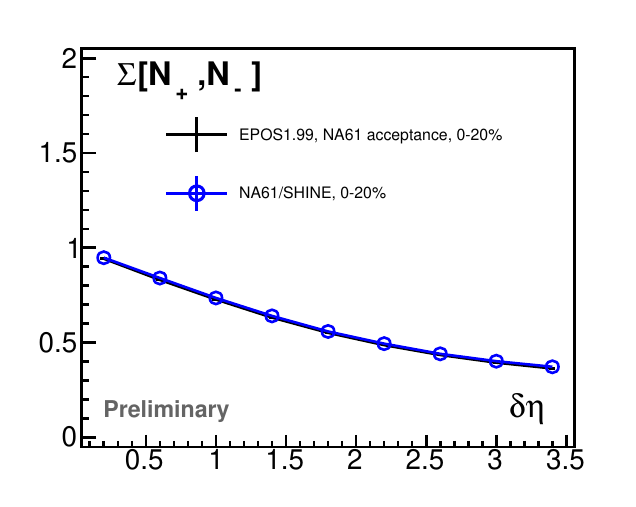}
\begin{minipage}[b]{14pc}\caption{\label{epos}$\Delta[N_{+},N_{-}]$ (left) and $\Sigma[N_{+},N_{-}]$ (right) as a function of the width of pseudorapidity interval $\delta\eta$ for 0-20\% centrality class (black) in comparison with predictions from the EPOS model (blue) for Be+Be interactions at $\sqrt{s_{NN}}=$ 16.83 GeV.}
\end{minipage}
\end{figure}

In Figure~\ref{epos} a comparison of $\Delta[N_{+},N_{-}]$ and $\Sigma[N_{+},N_{-}]$ for the 0-20\% centrality class with the prediction from the EPOS1.99 model~\cite{Werner:2007} is shown. Theoretical predictions were restricted to the NA61/SHINE geometrical acceptance. One can see that the behaviour of the measured charge fluctuation quantities is well reproduced by the EPOS model.
\ack
This work was supported by
the Federal Agency of Education of the Ministry of Education and Science of the
Russian Federation, SPbSU research grant 11.38.193.2014,
the Hungarian Scientific Research Fund (grants OTKA 68506 and 71989),
the J\'anos Bolyai Research Scholarship of
the Hungarian Academy of Sciences,
the Polish Ministry of Science and Higher Education (grants 667\slash N-CERN\slash2010\slash0, NN\,202\,48\,4339 and NN\,202\,23\,1837),
the Polish National Center for Science (grants~2011\slash03\slash N\slash ST2\slash03691, 2012\slash04\slash M\slash ST2\slash00816 and 
2013\slash11\slash N\slash ST2\slash03879),
the Foundation for Polish Science --- MPD program, co-financed by the European Union within the European Regional Development Fund,
the Russian Academy of Science and the Russian Foundation for Basic Research (grants 08-02-00018, 09-02-00664 and 12-02-91503-CERN),
the Ministry of Education, Culture, Sports, Science and Tech\-no\-lo\-gy, Japan, Grant-in-Aid for Sci\-en\-ti\-fic Research (grants 18071005, 19034011, 19740162, 20740160 and 20039012),
the German Research Foundation (grant GA\,1480/2-2),
the EU-funded Marie Curie Outgoing Fellowship,
Grant PIOF-GA-2013-624803,
the Bulgarian Nuclear Regulatory Agency and the Joint Institute for
Nuclear Research, Dubna (bilateral contract No. 4418-1-15\slash 17),
Ministry of Education and Science of the Republic of Serbia (grant OI171002),
Swiss Nationalfonds Foundation (grant 200020\-117913/1)
and ETH Research Grant TH-01\,07-3.


\section*{References}
\begin{thebibliography}{99}
\bibitem{Abgrall:2014fa} Abgrall N {\it et al.} (NA61 Coll.) 2014 {\it JINST} {\bf 9} P06005
\bibitem{Gazdzicki:1998vd} Gazdzicki M and Gorenstein M I 1999 {\it Acta Phys. Polon.} B {\bf 30} 2705 
\bibitem{Alt:2007aa} Alt C {\it et al.} (NA49 Coll.) 2008 {\it Phys. Rev.} C {\bf 77} 024903
\bibitem{Fodor:2004nz} Fodor Z and Katz S D 2004 {\it  JHEP} {\bf 0404} 050
\bibitem{Stephanov:1999zu} Stephanov M A, Rajagopal K and Shuryak E V 1999 {\it Phys. Rev.} D {\bf 60} 114028
\bibitem{Afanasiev:1999} Afanasiev S {\it et al.} (NA49 Coll.) 1999 {\it Nucl. Instrum. Meth.} A {\bf 430} 210
\bibitem{Grebieszkow:2009jr} Grebieszkow K for the NA49 Coll. 2009 {\it Nucl. Phys.} A {\bf 830} 546c–50c
\bibitem{Bialas:1976ed} Bialas A, Bleszynski M and Czyz W 1976 {\it Nucl. Phys.} B {\bf 111} 461
\bibitem{Gorenstein:2011vq} Gorenstein M I and Gazdzicki M 2011 {\it Phys. Rev.} C {\bf 84} 014904
\bibitem{Gazdzicki:2013ana} Gazdzicki M, Gorenstein M I and Mackowiak-Pawlowska M 2013 {\it Phys. Rev.} C {\bf 88} 024907
\bibitem{Czopowich:2014cp} Czopowich T. for the NA61/SHINE Coll. 2014 Transverse momentum and multiplicity fluctuations in the Be+Be energy scan from NA61/SHINE {\it Proceedings of Science} PoS(CPOD2014)054
\bibitem{Gazdzicki:2011xz} Gazdzicki M, Grebieszkow K, Mackowiak M and Mrowczynski S 2011 {\it Phys. Rev.} C {\bf 83} 054907 
\bibitem{Mackowiak-Pawlowska:2014ipa} Mackowiak-Pawlowska M and Wilczek A for the NA61 Coll. 2014 {\it J. Phys.: Conf. Series} {\bf 509} 012044
\bibitem{Jeon:2000} Jeon S and Koch V 2000 {\it Phys. Rev. Lett.} {\bf 85} 2076
\bibitem{Werner:2007} Werner K 2007 {\it Phys. Rev. Lett.} {\bf 98} 152301
\end{thebibliography}
\end{document}